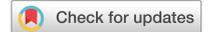

# OPEN  Twin-core fiber sensor integrated in laser cavity

Josu Amorebieta[1✉], Joao Pereira[2], Gaizka Durana[1], Carolina Franciscangelis[2], Angel Ortega-Gomez[1], Joseba Zubia[1], Joel Villatoro[1,3] & Walter Margulis[2,4]

In this work, we report on a twin-core fiber sensor system that provides improved spectral efficiency, allows for multiplexing and gives low level of crosstalk. Pieces of the referred strongly coupled multicore fiber are used as sensors in a laser cavity incorporating a pulsed semiconductor optical amplifier (SOA). Each sensor has its unique cavity length and can be addressed individually by electrically matching the periodic gating of the SOA to the sensor's cavity roundtrip time. The interrogator acts as a laser and provides a narrow spectrum with high signal-to-noise ratio. Furthermore, it allows distinguishing the response of individual sensors even in the case of overlapping spectra. Potentially, the number of interrogated sensors can be increased significantly, which is an appealing feature for multipoint sensing.

The use of optical fibers in sensing applications for real-time monitoring of parameters such as strain and temperature has attracted much interest, as in those fields, their intrinsic properties such as small size, lightweight and electromagnetic immunity can be exploited. Moreover, thanks to their capability to be embedded in materials such as concrete or composites, and to operate over long distances, they are an appealing alternative for many applications that require a precise tracking of any of the aforementioned parameters along large structures or areas[1]. To this end, multipoint sensing is often used, which consists of interrogating in a simple and versatile way several individual sensors[2,3], and whose spatial resolution is linked to the capability to discern between adjacent sensing elements. This configuration has gained much relevance for structure health monitoring in particular[4,5].

Among multipoint sensing techniques, the most common is wavelength division multiplexing[6], where each sensor operates at a different wavelength. Thus, the wavelength shift of each sensor and the interrogation window are the constraining factors that define the maximum number of elements that can be interrogated. In contrast, time division multiplexing is based on interrogating each sensing element individually by means of analyzing the reflected light[7], as the arrival times of the reflections are directly proportional to the distance from the light source to each sensing element. Furthermore, wavelength and time division multiplexing can be combined in order to increase the number of sensing elements that can be interrogated individually[8,9]. Commonly, the aforementioned techniques are implemented with fiber Bragg gratings (FBGs)[10–15], a mature and reliable technology for the measurement of multiple parameters[16] with a spatial resolution down to a few millimeters[17]. Moreover, FBGs make an efficient use of the spectrum, as they provide narrow and well-defined peaks. This fact allows monitoring a significant amount of FBGs in the same interrogation window. As an alternative, in-fiber Fabry-Perot[18,19] and Mach-Zehnder[20] interferometers are used as well.

In recent years, strongly coupled multicore fibers (MCFs) have been introduced as an alternative for sensing. Some appealing features of MCFs are their versatility, interrogation ease, sensitivity[21–24], which can be higher than those of FBGs depending on the measurand[25], and that km-lengths of fiber can be manufactured, leading to the availability from one single draw of many thousands of decimeter-long fiber segments appropriate for sensing. Their main drawback is their low spectral efficiency, as coupled MCFs provide multiple and broad peaks. The spectral overlap of the broad peaks of C-band sensors generally constrains their use to single point measurements. To overcome this limitation, efforts such as cascading MCF segments have been made, at the expense of increasing the complexity and length of the sensor[26,27]. Anyhow, due to the attractive features of coupled MCF-based sensors, it would be of interest to develop a technique to improve their spectral efficiency and allow for multiplexing, making it possible their use as multipoint sensors for applications such as structural health monitoring. Among MCFs, the use of twin core fibers (TCFs) as sensing elements for multiple applications has been widely reported in the literature[28–31].

[1]Department of Communications Engineering, University of the Basque Country UPV/EHU, 48013 Bilbao, Spain. [2]Fiber Optics, RISE Research Institutes of Sweden, 164 40 Stockholm, Sweden. [3]Ikerbasque-Basque Foundation for Science, 48011 Bilbao, Spain. [4]Department of Applied Physics, Royal Institute of Technology, 106 91 Stockholm, Sweden. ✉email: josu.amorebieta@ehu.eus

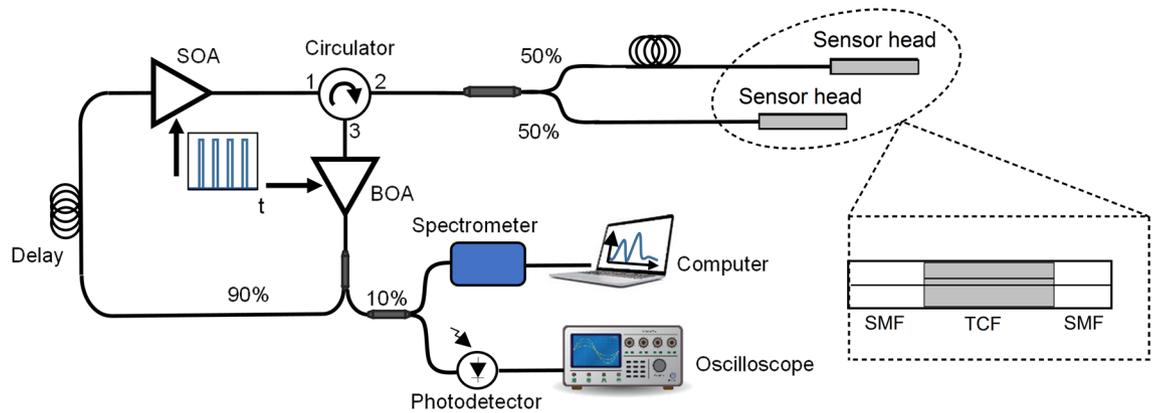

**Figure 1.** Schematic of the interrogation setup.

In this paper, a multipoint interrogation system with MCFs is presented. It is based on using segments of a coupled twin-core fiber (TCF) as sensing elements inside a laser cavity that incorporates a pulsed semiconductor optical amplifier (SOA). As explained and demonstrated in the following sections, lasing improves the spectral efficiency of the sensor, and the system allows for multiplexing, increasing the number of sensors that can be interrogated unmistakably.

## Operational mechanism and experimental setup

In the proposed system, the key to multiplex several MCF-based sensors relies on creating ring cavities with different lengths that resonate at a particular RF frequency in the MHz range. In each cavity, the sensing device acts as a laser mirror that reflects the light that is launched by a SOA. The latter is gated with a waveform generator, which allows programming the frequency rate and width of the emitted pulse. As a result, only the optical feedback from a sensor head that is synchronous to the nanosecond electrical gating of the SOA is amplified and recirculates through the cavity, whereas the optical signal arriving at another time is absorbed. When the optical pulse passes multiple times through the SOA, and provided the amplification exceeds the loss, the interrogator operates as a laser, whose emission takes place at the cavity gain peak with significant spectral narrowing[17,32,33], good signal-to-noise ratio (SNR) and low level of crosstalk. Moreover, as each one of the sensing elements is addressed exclusively at its own resonance frequency ($f_r$), sensors with identical or similar spectra can be used if deployed at different lengths[34,35].

The setup of the proposed system is shown in Fig. 1. It consists of two polarization-independent semiconductor optical amplifiers (SOA, Thorlabs SOA1013SXS and Booster Optical Amplifier BOA, Thorlabs BOA1004P), a 90:10 fiber optic coupler for monitoring, a single mode fiber reel (SMF), a circulator and a 50:50 coupler to access the sensors. The use of two amplifiers (SOA and BOA) in the laser cavity was required to increase the gain in order to ensure enough amplification for the system to lase. Both amplifiers were gated at the same rate and with an appropriate phase shift by a waveform generator (Keysight 33600A) that triggered the high-speed current amplifiers to drive the SOA and BOA. The overall pulse duration when lasing was below 10 ns, i.e., covering 1 m of fiber in reflection. The 320 m reel of SMF was used for convenience to increase the cavity roundtrip time and lower the gating frequency to values compatible with the available electronics, avoiding excessive heating. The relatively long fiber reel also ensured that any of the lengths of the cavities formed by the sensors were not multiple integers of each other[32]. Port 2 of the circulator was used to connect the sensors to the ring by means of a 50:50 coupler. The sensors were located at different lengths in order to ensure a different $f_r$ for each so that they could be addressed individually by sweeping the RF frequency of the current pulses to the SOA. Mathematically, the $f_r$ of each device can be calculated with the following equation[33,35]:

$$f_r = \frac{c}{n(L_{RING} + 2L_{SENSOR})}, \quad (1)$$

where $c$ is the speed of light, $n$ is the effective refractive index of the probed fiber, $L_{RING}$ is the cavity length of the ring formed by the SMF reel, the SOA, the BOA and Ports 1 and 3 of the circulator, which was fixed to 340 m in this work, and $L_{SENSOR}$ is the distance from Port 2 to the end of each of the outputs of the fiber optic coupler. Typically, the resonance frequency is in the $\sim 10^6$ Hz range (for a sensor distance of $L \sim 10^2$ m), and variations due to temperature fluctuations in a laboratory are limited to $\sim 10^1$ Hz (i.e., an error in the sensor position in the mm range). For the setup in Fig. 1, $f_{rSENSOR1}$ and $f_{rSENSOR2}$ were found to be 537.16 kHz and 544.49 kHz, respectively. Outside those repetition frequencies there was no measurable optical signal in the interrogation window, as the laser was below threshold except when at resonance.

For monitoring purposes 10% of the light was used: half connected to a spectrometer (I-MON512-USB, Ibsen Photonics) and the other half to an InGaAs avalanche photodetector (Thorlabs APD430C/M) and an oscilloscope (Tektronix TDS3034) for spectral and temporal measurements, respectively. With the former, any wavelength shift in the spectrum could be tracked and stored with picometer accuracy, whereas with the latter, any shortening of the optical pulse could be measured, as the traces of electric pump and optical pulses could be displayed.

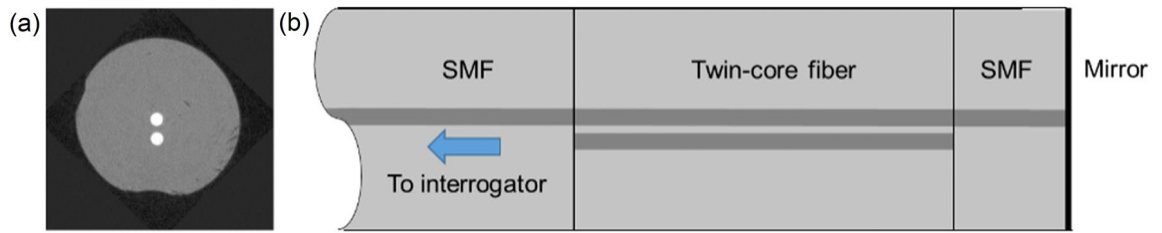

**Figure 2.** (**a**) Cross-section of the 125 μm TCF used in this work. (**b**) Schematic of the sensor head comprising a section of TCF and a short piece of mirrored single-mode fiber.

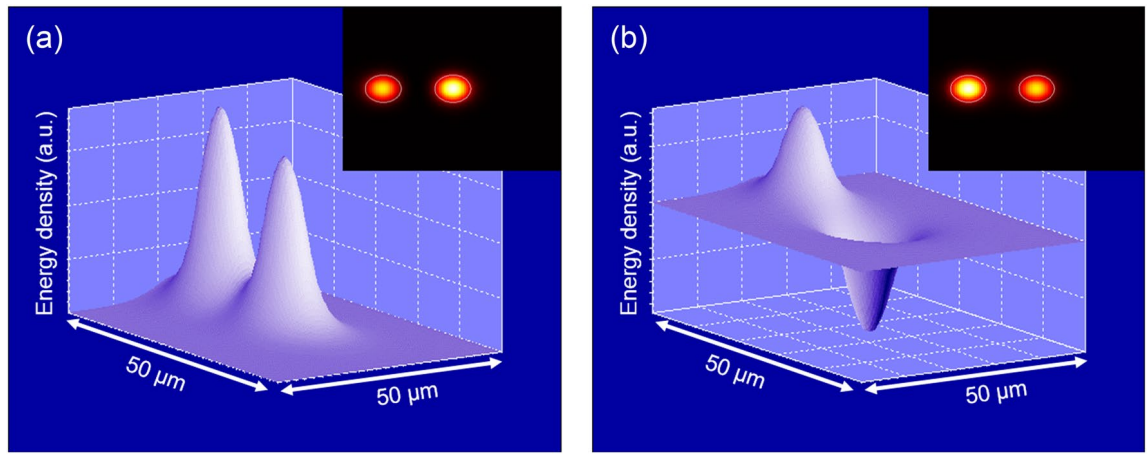

**Figure 3.** Simulations of the 3D and 2D profiles of the two coupled orthogonal supermodes excited in the TCF. In (**a**) the supermode $SP_{01}$ is shown, and in (**b**) the supermode $SP_{02}$.

Such interrogation system has been combined typically with FBGs[17,32–35]. However, in this work, the novelty relies on the fact that sections of coupled TCF were used as sensing elements and partial laser cavity mirrors, as combining the proposed interrogation system and coupled MCFs could merge the benefits of both elements. On the one hand, the interrogation system allows for multiplexing, optimizing the spectral efficiency of the sensors, interrogating them individually and using devices with similar or even identical spectra, as they are identified by their $f_r$ and not by the shape of their spectra. On the other hand, MCFs provide several intrinsic advantages such as high sensitivity, ease of interrogation, possibility to design ad-hoc geometries to optimize their sensitivity for the measurement of certain parameters[36,37], etc. Thus, the potential benefits of the combination of the proposed interrogation system with MCFs are of big interest for multipoint sensing systems.

The TCF used in this work was manufactured from a standard telecom preform with a central core and one extra core added to its drilled cladding. The two cores were approximately equal in size and physical properties, and the center of the side-core was located 15.5 μm away from the fiber center (see Fig. 2a). The mean diameter and numerical aperture (NA) of each of the cores were 8.2 μm and 0.14, respectively, to match those of a standard SMF, while the mean diameter of the fiber was 125 μm[28].

The operating principle of coupled MCFs in general and of this TCF in particular can be explained by the Coupled Mode Theory, as it can be assumed that each of the cores acts as a waveguide[38]. Its particularization for optical fibers and its corresponding theoretical and mathematical analysis is explained in detail in Refs.[39,40]. Briefly, the supermodes that propagate in coupled MCFs are the linear combination of the propagating modes through each individual guide[41]. When the TCF is fusion spliced to an SMF with radial symmetry and excited with the $LP_{01}$ mode, only the two orthogonal supermodes that have power in the central core will be coupled. For the TCF used in this work (see Fig. 2a), only the two coupled orthogonal propagating supermodes $SP_{01}$ and $SP_{02}$ are excited, as shown in Fig. 3.

According to the mathematical expressions in Refs.[39,40], the power transfer is cyclical among the cores of the fiber, and depends on the launched wavelength $\lambda$, the length of the TCF segment $z$ and the difference between the effective refractive indices of the propagating orthogonal supermodes $\Delta n$. Here, the normalized powers in the central ($P_0$) and adjacent core ($P_1$) are given by[28]:

$$P_0(z) = \cos^2\left(\frac{\pi \Delta n}{\lambda} z\right), \tag{2}$$

$$P_1(z) = \sin^2\left(\frac{\pi \Delta n}{\lambda} z\right), \tag{3}$$

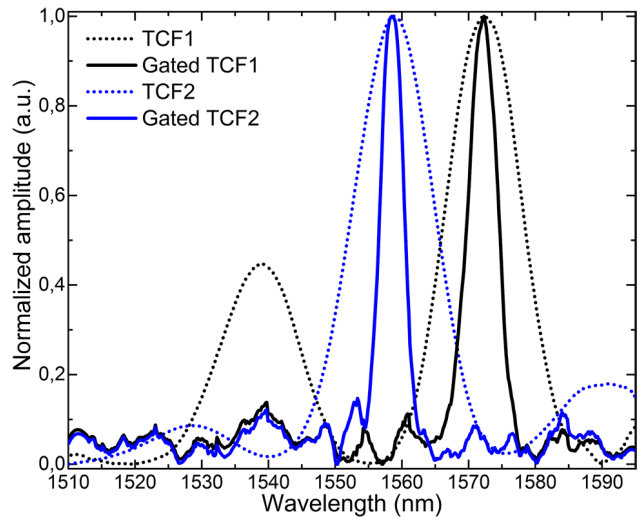

**Figure 4.** Spectra of the manufactured TCF sensors when they are probed by a broadband light source in a conventional open loop configuration (dotted line) and as part of a pulsed laser cavity, tuning the repetition frequency to the resonant cavity length of each (continuous line).

where $\Delta n = 0.873 \times 10^{-4}$ for this TCF at 1550 nm and at room temperature according to the results provided by PhotonDesign simulation software. Equations (2) and (3) highlight the fact that the optical power is transferred fully from one core to the other periodically. In the SMF-TCF-SMF structure shown in Fig. 2b, which was used as sensor head in this work, when a broadband light is launched into the central core of the TCF and analyzed at its exit by a spectrometer, a series of maxima and minima appear in the spectrum. However, when the same SMF-TCF-SMF structure is interrogated in reflection[21,27,36,37], the light travels back and forth through it and the normalized coupled power at the output becomes:

$$P_0(z) = \cos^4\left(\frac{\pi \Delta n}{\lambda} z\right), \quad (4)$$

$$P_1(z) = \sin^4\left(\frac{\pi \Delta n}{\lambda} z\right). \quad (5)$$

As it can be noticed from Eqs. (2) to (5), the coupling period is unaltered regardless of interrogating the structure in transmission or reflection. The wavelength at which the maxima takes place in the spectrum ($\lambda_m$) is unaltered as well, and occurs when the period equals a multiple integer (*m*) of π.

$$\lambda_m = \frac{\Delta n z}{m}. \quad (6)$$

Thus, if we excite the TCF with a SOA whose pulses are tuned to the $f_r$ of the cavity (~ $10^6$ Hz), the system will lase at $\lambda_m$, providing a unique, much sharper and narrower peak in the spectrum compared to the broad maxima and minima obtained when such MCFs are interrogated with a broadband light source[21–23]. This fact allows for multiple-peak clean-up, and using and multiplexing MCF-based devices with identical or similar spectra in the same interrogation window, as they are unmistakably and individually identified by the unique $f_r$ of each. Moreover, if the TCF is exposed to an external perturbation such as a vibration or strain, its RF resonance frequency will not change whereas its lasing wavelength $\lambda_m$ will shift according to the perturbation. Such shift can be correlated with the perturbation; hence, the referred system can be used for measurement and sensing purposes.

To demonstrate such operating principle, two SMF-TCF-SMF structures were manufactured (TCF1 and TCF2) and placed in the setup in Fig. 1 as sensing devices, each of them with a TCF segment of ~ 8 cm but with the segment in TCF1 slightly longer than that in TCF2 ($L_{TCF1} > L_{TCF2}$). To that end, a precision fiber cleaver and a precision fusion splicer were used. With the former, fiber segments with cleaving angles close to 0° were obtained, whereas the latter was used to align and splice the central core of the TCF with that of the SMF with low insertion losses. For comparison purposes, both samples were interrogated by a continuous broadband light source (dashed trace) and in a laser cavity (solid trace), as shown in Fig. 4.

As predicted, the spectra of the devices that are interrogated via intracavity in a lasing configuration are much narrower (FWHM = 5.13 nm for TCF1 and FWHM = 4.26 nm for TCF2) than those from the samples interrogated with a continuous broadband light source and in an open loop (FWHM = 12.83 nm for TCF1 and FWHM = 13.79 nm for TCF2). The optical signal measured in a lasing mode of operation consists of a single peak, in contrast to the response of the same sensor operated in the conventional configuration, when other peaks of comparable amplitude are visible, making it difficult to be used for sensing. This spectral narrowing and multiple-peak clean-up achieved with the proposed lasing configuration allows multiplexing sensors without peak superposition.

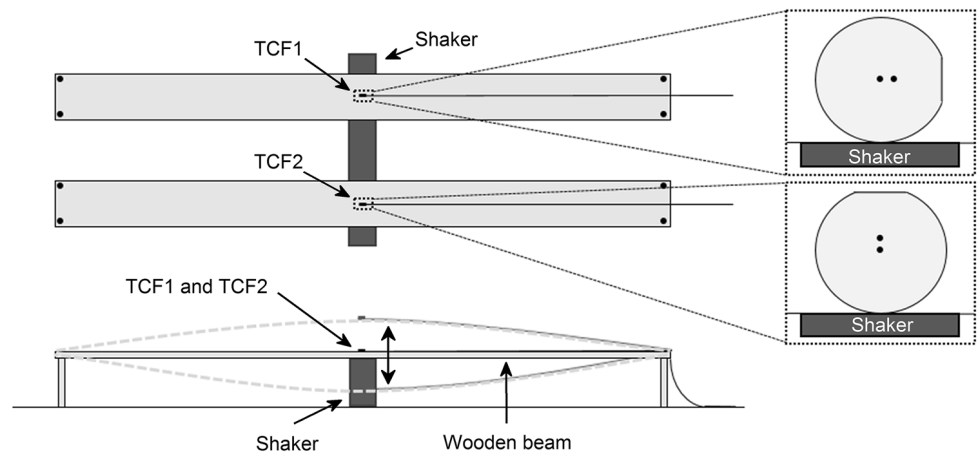

**Figure 5.** Schematic top and lateral views of the experimental setup. The lateral view shows how the vibration was applied. The close-up shows how the sensor heads were surface bonded to each beam.

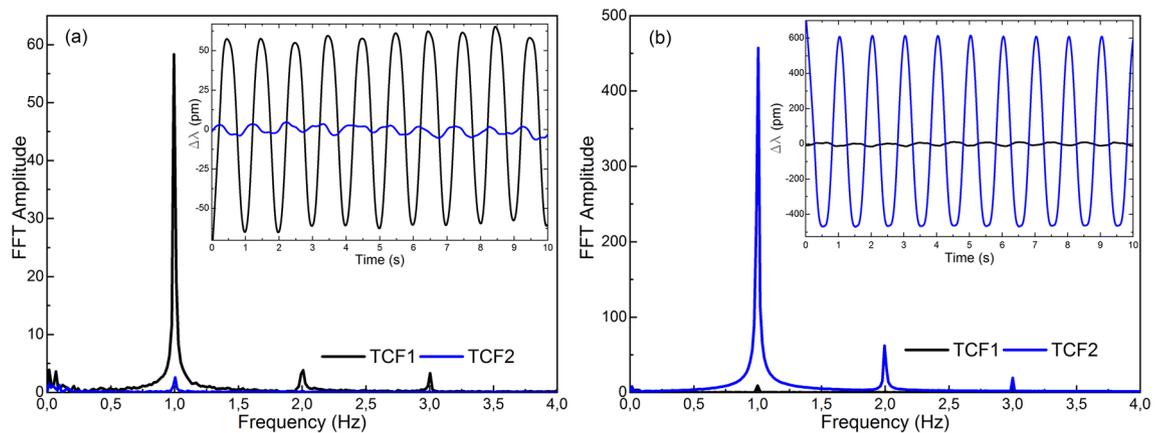

**Figure 6.** Results in frequency domain when (**a**) TCF1 was vibrating and TCF2 was inactive, and (**b**) TCF1 was inactive and TCF2 was vibrating. In the insets, the results in time domain of each case are shown.

## Using the sensors: results and discussion

The first test consisted in subjecting both samples to vibrations of identical frequency (1 Hz) and amplitude (500 mVpp). To that end, TCF1 and TCF2 were surface bonded longitudinally to the center of the upper side of two thin and flexible wooden beams of ~ 1 m length. As illustrated in Fig. 5, the beams were parallel and supported at the extremes, while the vibrations were applied to the center of each beam by a shaker connected to a function generator (Hewlett Packard 33120A) and a 4Ω output impedance audio amplifier (Brüel and Kjaer). With this setup, it was possible to subject to vibrations only one or both beams at the same time, which permitted evaluating the crosstalk. TCF1 and TCF2 were surface bonded to each beam with their cores oriented perpendicularly to each other: TCF1 had its cores oriented horizontally and TCF2 vertically, respectively. To that end, the process described in Ref.[21] was replicated. The sensitivity of asymmetric MCFs as the one used in this work depends on the relation between the orientation of the cores and the plane in which the effect is applied[21,25,42,43]. Thus, as vibrations were applied vertically, regarding the expected wavelength shift, TCF1 would have minimum sensitivity, whereas TCF2 would have maximum sensitivity. Moreover, the flat surface in the TCF made the alignment process easier, as it is aligned perpendicular to the cores (see Fig. 2a). Thus, just by pressing the fiber against a flat surface, the direction of the cores could be known. To create the flat surface, the preform was mounted on a lathe and rotated until the cores were vertical. After that, the preform was flattened on the top-most surface.

The results are summarized in Figs. 6 and 7. In Fig. 6a the results of exposing TCF1 to vibrations and leaving TCF2 in idle are shown, in Fig. 6b TCF2 was subjected to vibrations whereas TCF1 was left inactive, and in Fig. 7 both sensors were subjected to vibrations at the same time. In each case, both sensors were interrogated by adjusting the frequency of the gating to the resonance frequency of the cavity.

For the cases in which only one of the sensors was vibrating, the good SNR measured at 1 Hz, the narrowness of the peak and the low level of the harmonic components in the frequency domain were significant. In fact, the SNRs for the cases in Fig. 6a,b were 21.74 and 47.13, respectively. Concerning the difference in wavelength shift between sensors against the same perturbation, it was caused by the core orientation. As expected, as TCF2 was aimed at maximizing the sensitivity, the wavelength shift was much higher in this device. Regarding the sensor that was resting in each case, results indicate that the interference in the signal caused by the sensor that was vibrating was low. In the time domain, this fact is reflected in the small wavelength shift compared to that of the vibrating sensor. In Fig. 6a, the shift recorded in the resting sensor is only 4.6% of the signal of the vibrating sensor, whilst in Fig. 6b, it is of 2.1%. In the frequency domain, the amplitude of the FFT peak in 1 Hz is less than an order of magnitude below that of the sensor that is vibrating.

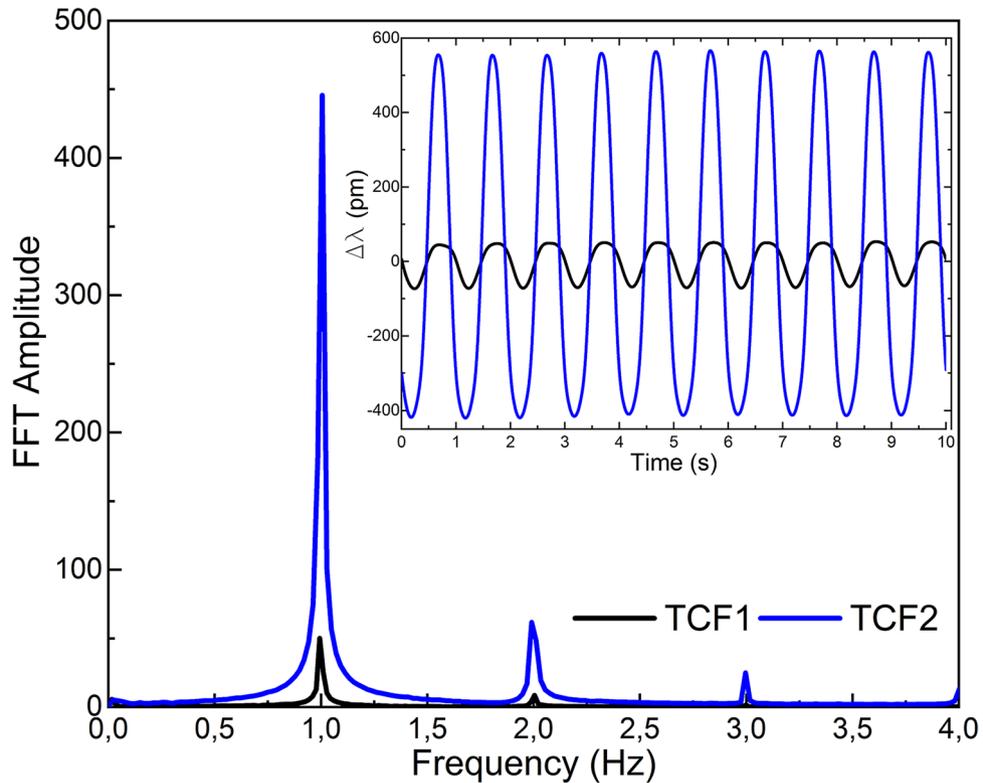

**Figure 7.** Results in frequency domain when both devices were vibrating. Results in time domain are shown in the inset.

When both sensors were subjected to identical vibrations simultaneously (see Fig. 7), TCF2 showed higher sensitivity than TCF1. The latter was expected due to the already described perpendicular orientation of the cores (see Fig. 5). In the time domain, the wavelength shift of TCF2 was approximately an order of magnitude higher than that in TCF1 ($\Delta\lambda_{TCF2}$ = 1060 pm and $\Delta\lambda_{TCF1}$ = 142 pm). In the frequency domain, a peak in 1 Hz and low level of harmonic components in the signal from both sensors was noticeable. Thus, the vibration was detected and measured with high accuracy by both sensors and the crosstalk had no significant effect on the measurements nor on their sensitivity.

The next test consisted in evaluating the performance of the interrogation system when more than two sensors were connected at the same time. To that end, the setup in Fig. 1 was modified to include a variable attenuator (Hewlett Packard 8157A) between Port 2 of the circulator and the 50:50 coupler. It was set to 6 dB, which, added to the 3 dB losses of the 50:50 coupler, generated a total loss of 9 dB in each output in order to simulate the losses of a system in a configuration comprising as many as 8 sensors. As for the previous test, both sensors were subjected to vibrations (1 Hz and 500 mVpp) and interrogated by adjusting the frequency of the gating to the resonance frequency of each. The addition of the attenuator increased slightly the length of the fiber cavity and therefore, the resonance frequency of the sensors changed. For this case, they were 328.3 kHz and 353.4 kHz for TCF1 and TCF2, respectively.

The results are summarized in Figs. 8 and 9. In Fig. 8a the results of exposing TCF1 to vibrations and leaving TCF2 in idle are shown, in Fig. 8b TCF2 was subjected to vibrations whereas TCF1 was left inactive, and in Fig. 9 both sensors were subjected to vibrations at the same time.

The 1 Hz oscillation was recorded by the device that was subjected to vibrations in each case (TCF1 in Fig. 8a and TCF2 in Fig. 8b), whereas the signal from the resting device did not detect any oscillation. In fact, the FFT of each case shows that the SNR is always above 3, which is the ratio commonly taken as a rule to define the limit of detection[44]. Thus, results indicate that both sensors were able to detect and record the 1 Hz vibration with a level of interference that did not affect their performance.

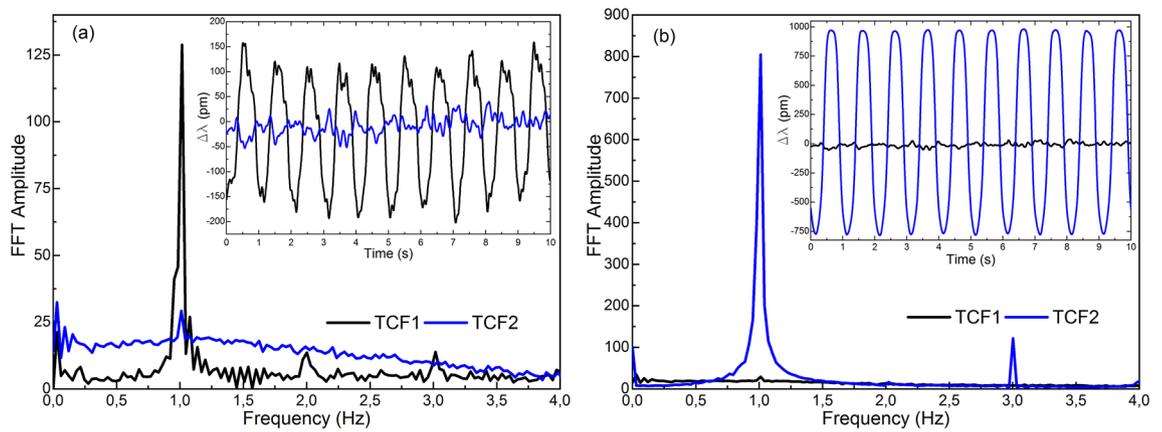

**Figure 8.** Results in frequency domain when (**a**) TCF1 was vibrating and TCF2 was inactive and (**b**) TCF1 was inactive and TCF2 was vibrating. In the insets, the results in time domain of each case are shown.

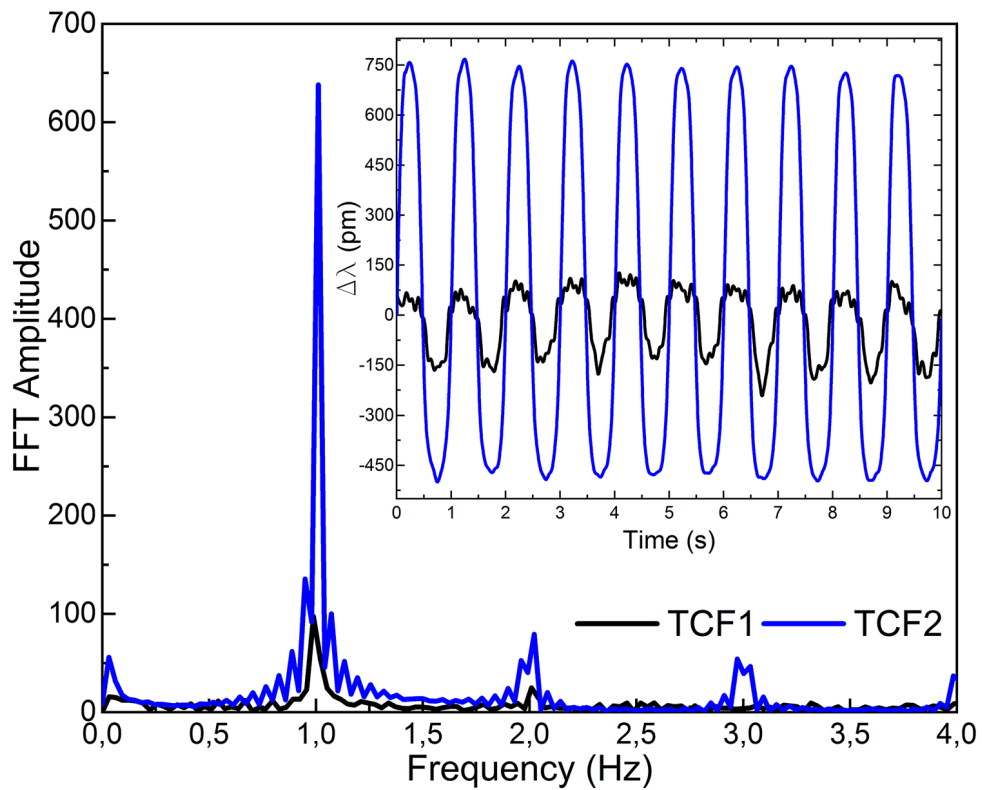

**Figure 9.** Results in frequency domain when both devices were vibrating. Results in the time domain are shown in the inset.

When the vibration was applied simultaneously to both devices (see Fig. 9), the oscillation was recorded by both, but with slightly lower sensitivities. As it happened in the previous tests, in all the cases, TCF2 showed higher sensitivity due to its better core orientation.

The last test consisted in measuring different parameters with each sensing head. Taking into consideration that the temperature sensitivity of this TCF has been reported already[28] (46.4 pm/K), and that the aim of this work is to propose and demonstrate a multiplexing system for MCFs, it was decided to measure the performance of the TCF against two parameters different from temperature. These parameters were selected to be sinusoidal vibrations (1 Hz and 500 mVpp), and strain cycles consisting of stretching and compression cycles in steps of 257.57 μm by means of a linear precision stage (Newport M-UMR 12.40). To that end, the attenuator was removed so that the experimental setup for these tests was identical to that in Fig. 1. TCF2 was chosen to be used for vibration measurements, as it was already glued to a wooden beam and had a better core orientation compared with TCF1 (see Fig. 5). The latter was replaced by TCF3 (see Fig. 10), a sample with a slightly longer TCF segment (~ 12 cm) and whose $f_{rSENSOR3}$ ~ 566.9 kHz.

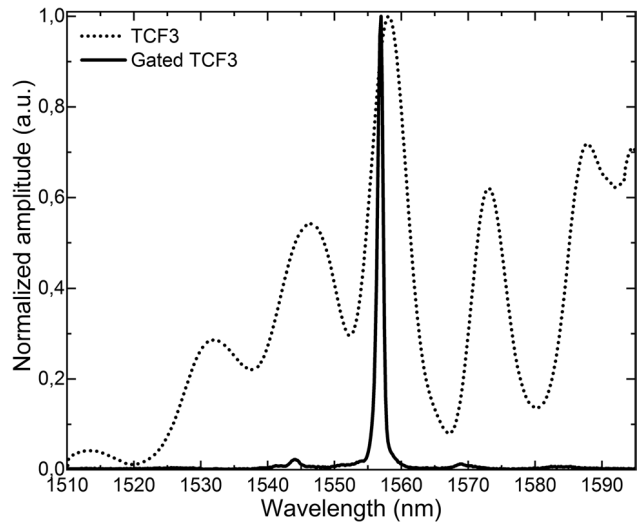

**Figure 10.** Spectra of the manufactured TCF3 when it is probed by a broadband light source in a conventional open loop configuration (dotted line) and as part of a pulsed laser cavity, tuning the repetition frequency to the resonant cavity length (continuous line).

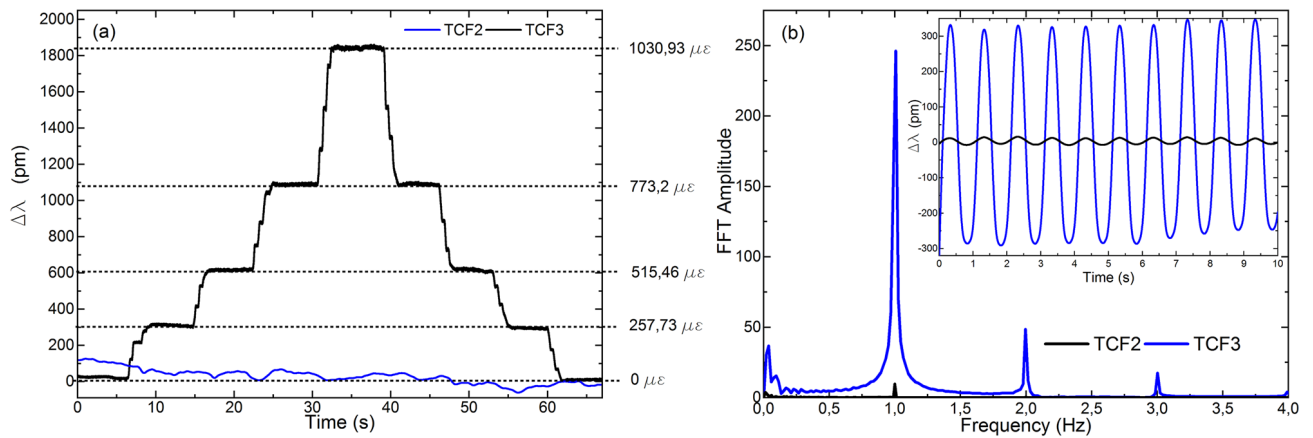

**Figure 11.** Results when (**a**) TCF3 was subjected to strain and TCF2 was inactive, and (**b**) TCF3 was inactive and TCF2 was vibrating. In the inset, the result in the time domain is shown.

When TCF3 was interrogated with a continuous broadband light source, the FWHM of the maximum peak was of 6.63 nm, whereas the FWHM of the unique peak that appears on the spectrum when it was interrogated by the pulsed SOA was only of 1.16 nm.

The results from these tests are shown in Figs. 11 and 12. In Fig. 11a, TCF3 was subjected to strain whereas TCF2 was left in idle state, in Fig. 11b TCF2 was subjected to vibrations and TCF3 was left inactive, and in Fig. 12 TCF2 and TCF3 were subjected to vibrations and strain, respectively and simultaneously.

Results indicate that vibration and strain cycles have been tracked correctly by each sensor in all the tests. On the one hand, for strain measurements, when only TCF3 was subjected to it, the SNR was of 27.38 and the wavelength shift was of 1875 pm in the tested range (see Fig. 11a), which indicates that TCF3 reached a sensitivity of 1.8 pm/με. When both sensors were tested (see Fig. 12), TCF3 showed a sensitivity of 1.7 pm/με, as the wavelength shift was of 1750 pm in the tested range. These results indicate that this TCF has a significant performance for strain measurements, as its sensitivity is higher than that reported by other MCF-based strain sensors[23], and that the crosstalk caused a sensitivity loss of 6.7%. On the other hand, when only TCF2 was vibrating, the SNR was of 34.32 and the wavelength shift was of 600 pm (see Fig. 11b), whereas it was of 450 pm when both sensors were tested (see Fig. 12). Thus, for this case, the loss in sensitivity caused by crosstalk was of 25%. In any case, the effect of the latter did not have a significant impact on the measurements regardless of the measurand, as the shift caused by both parameters was recorded accurately.

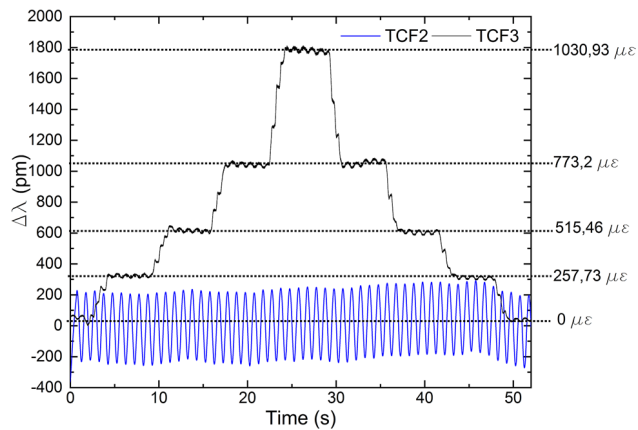

**Figure 12.** Results when TCF2 was vibrating and TCF3 was strained simultaneously.

### Conclusions
In this work, we have proposed and demonstrated the feasibility of a multipoint interrogation system for coupled MCF-based sensors. Its operating principle is based on creating several ring resonant cavities in which each of the sensor heads acts as a laser mirror that reflects the light that is launched by a gated SOA. In this manner, each cavity can be easily and individually addressed by tuning the frequency of the gating, as the latter is directly proportional to the physical length of each of the cavities. Thus, if the sensors are distributed at different lengths, each of them will have a unique resonant frequency that will allow for their unmistakable identification. Moreover, as the identification depends solely on the length of the cavity, sensors with similar or even identical spectra could be multiplexed in the same system.

A major advantage provided by the proposed system is the significant improvement of the spectral efficiency of coupled MCF-based sensors. Compared to the output spectra provided by these devices, which is comprised of multiple and broad peaks when they are interrogated by a continuous broadband light source; when the same devices are interrogated by the gated SOA, the output spectra from each sensor solely has a unique, sharp and narrow peak, as the system makes a multiple peak clean-up and only lases at the wavelength at which the cavity gain peak is located. These features make it possible to include a larger number of sensors in the interrogation window, which added to the possibility of using sensors with identical spectra, allows increasing significantly the number of devices that could be interrogated by means of this system.

The feasibility of the proposed interrogation system has been demonstrated for a setup with 2 sensors that consisted of segments of TCF of different lengths that were subjected to the same (vibration) and to different stimuli (vibration and strain). In all cases, the system was able to detect and measure the applied effect with high sensitivity, high SNR and a significantly low level of crosstalk, even in a configuration that simulated a situation with 8 sensors.

Hence, the high sensitivity provided by MCFs for the simultaneous measurement of different parameters, combined with the high spectral efficiency, high SNR and low level of crosstalk provided by the interrogation setup, suggest that the proposed system may be appealing for applications that require accurate multipoint sensing with minimal intrusiveness and interrogation ease, such as structural health monitoring of aircrafts, buildings, etc. In addition, as the proposed system is compatible with any MCF, it allows for multiplexing and interrogating MCFs with different geometries simultaneously within the same setup.

### Data availability
The datasets used and/or analysed during the current study available from the corresponding author on reasonable request.

### References
 1. Bado, M. F. & Casas, J. R. A review of recent distributed optical fiber sensors applications for civil engineering structural health monitoring. *Sensors*. https://doi.org/10.3390/s21051818 (2021).
 2. Li, W., Yuan, Y., Yang, J. & Yuan, L. Review of optical fiber sensor network technology based on white light interferometry. *Photonic Sens.* **11**, 31–44. https://doi.org/10.1007/s13320-021-0613-x (2021).
 3. Senior, J. M., Moss, S. E. & Cuswo, D. Multiplexing techniques for noninterferometric optical point-sensor networks: A review. *Fiber Integr. Opt.* **17**, 3–20. https://doi.org/10.1080/014680398245028 (1998).
 4. Lorenzo, J. & Doll, W. (eds) *Levees and Dams: Advances in Geophysical Monitoring and Characterization* 91–120 (Springer, 2019).
 5. Maaskant, R. *et al.* Fiber-optic Bragg grating sensors for bridge monitoring. *Cement Concr. Compos.* **19**, 21–33. https://doi.org/10.1016/S0958-9465(96)00040-6 (1997).
 6. Senior, J. M. & Cusworth, S. D. Wavelength division multiplexing in optical fibre sensor systems and networks: A review. *Opt. Laser Technol.* **22**, 113–126. https://doi.org/10.1016/0030-3992(90)90021-U (1990).

7. Cooper, D. J. F., Coroy, T. & Smith, P. W. E. Time-division multiplexing of large serial fiber-optic Bragg grating sensor arrays. *Appl. Opt.* **40**, 2643–2654. https://doi.org/10.1364/AO.40.002643 (2001).
8. Liu, D., Sun, Q., Lu, P., Xia, L. & Sima, C. Research progress in the key device and technology for fiber optic sensor network. *Photonic Sens.* **6**, 1–25 (2016).
9. Vohra, S., Dandridge, A., Danver, B. & Tveten, A. *Optical Fiber Sensors* (Optical Society of America, 1996).
10. Childs, P. An FBG sensing system utilizing both WDM and a novel harmonic division scheme. *J. Lightwave Technol.* **23**, 348 (2005).
11. Yan, L. *et al.* High-speed FBG-based fiber sensor networks for semidistributed strain measurements. *IEEE Photonics J.* **5**, 7200507. https://doi.org/10.1109/JPHOT.2013.2258143 (2013).
12. Luo, Y. *et al.* Optical chaos and hybrid WDM/TDM based large capacity quasi-distributed sensing network with real-time fiber fault monitoring. *Opt. Express* **23**, 2416–2423. https://doi.org/10.1364/OE.23.002416 (2015).
13. Valente, L. C. G. *et al.* In *2002 15th Optical Fiber Sensors Conference Technical Digest. OFS 2002 (Cat. No.02EX533)*, Vol. 151, 151–154.
14. Kashyap, R. *Fiber Bragg Gratings* (Academic Press, 2009).
15. Othonos, A., Kalli, K., Pureur, D. & Mugnier, A. *Wavelength Filters in Fibre Optics* 189–269 (Springer, 2006).
16. Chen, J., Liu, B. & Zhang, H. Review of fiber Bragg grating sensor technology. *Front. Optoelectron. China* **4**, 204–212 (2011).
17. Margulis, W., Lindberg, R., Laurell, F. & Hedin, G. Intracavity interrogation of an array of fiber Bragg gratings. *Opt. Express* **29**, 111–118. https://doi.org/10.1364/OE.414094 (2021).
18. Yoshino, T., Kurosawa, K., Itoh, K. & Ose, T. Fiber-optic Fabry-Perot interferometer and its sensor applications. *IEEE Trans. Microw. Theory Tech.* **30**, 1612–1621 (1982).
19. Rao, Y.-J., Ran, Z.-L. & Gong, Y. *Fiber-Optic Fabry-Perot Sensors: An Introduction* (CRC Press, 2017).
20. Zhao, Y., Zhao, H., Lv, R.-Q. & Zhao, J. Review of optical fiber Mach–Zehnder interferometers with micro-cavity fabricated by femtosecond laser and sensing applications. *Opt. Lasers Eng.* **117**, 7–20 (2019).
21. Amorebieta, J. *et al.* Compact omnidirectional multicore fiber-based vector bending sensor. *Sci. Rep.* **11**, 5989. https://doi.org/10.1038/s41598-021-85507-9 (2021).
22. Amorebieta, J. *et al.* Packaged multi-core fiber interferometer for high-temperature sensing. *J. Lightwave Technol.* **37**, 2328–2334. https://doi.org/10.1109/JLT.2019.2903595 (2019).
23. Villatoro, J. *et al.* Accurate strain sensing based on super-mode interference in strongly coupled multi-core optical fibres. *Sci. Rep.* **7**, 4451. https://doi.org/10.1038/s41598-017-04902-3 (2017).
24. Villatoro, J., Antonio-Lopez, E., Schülzgen, A. & Amezcua-Correa, R. Miniature multicore optical fiber vibration sensor. *Opt. Lett.* **42**, 2022–2025. https://doi.org/10.1364/OL.42.002022 (2017).
25. Arrizabalaga, O. *et al.* High-performance vector bending and orientation distinguishing curvature sensor based on asymmetric coupled multi-core fibre. *Sci. Rep.* **10**, 14058. https://doi.org/10.1038/s41598-020-70999-8 (2020).
26. Villatoro, J. *et al.* Composed multicore fiber structure for direction-sensitive curvature monitoring. *APL Photonics* **5**, 070801. https://doi.org/10.1063/1.5128285 (2020).
27. Amorebieta, J. *et al.* Highly sensitive multicore fiber accelerometer for low frequency vibration sensing. *Sci. Rep.* **10**, 16180. https://doi.org/10.1038/s41598-020-73178-x (2020).
28. Rugeland, P. & Margulis, W. Revisiting twin-core fiber sensors for high-temperature measurements. *Appl. Opt.* **51**, 6227–6232. https://doi.org/10.1364/AO.51.006227 (2012).
29. Flores-Bravo, J. A., Madrigal, J., Zubia, J., Sales, S. & Villatoro, J. Coupled-core fiber Bragg gratings for low-cost sensing. *Sci. Rep.* **12**, 1280. https://doi.org/10.1038/s41598-022-05313-9 (2022).
30. Wu, Y. *et al.* Highly sensitive curvature sensor based on asymmetrical twin core fiber and multimode fiber. *Opt. Laser Technol.* **92**, 74–79. https://doi.org/10.1016/j.optlastec.2017.01.007 (2017).
31. Li, Z. *et al.* Highly-sensitive gas pressure sensor using twin-core fiber based in-line Mach-Zehnder interferometer. *Opt. Express* **23**, 6673–6678. https://doi.org/10.1364/OE.23.006673 (2015).
32. Lloyd, G. D., Everall, L. A., Sugden, K. & Bennion, I. Resonant cavity time-division-multiplexed fiber Bragg grating sensor interrogator. *IEEE Photonics Technol. Lett.* **16**, 2323–2325. https://doi.org/10.1109/LPT.2004.834849 (2004).
33. Kersey, A. D. & Morey, W. W. Multiplexed Bragg grating fibre-laser strain-sensor system with mode-locked interrogation. *Electron. Lett.* **29**, 112–114. https://doi.org/10.1049/el:19930073 (1993).
34. Dai, Y., Liu, Y., Leng, J., Deng, G. & Asundi, A. A novel time-division multiplexing fiber Bragg grating sensor interrogator for structural health monitoring. *Opt. Lasers Eng.* **47**, 1028–1033. https://doi.org/10.1016/j.optlaseng.2009.05.012 (2009).
35. Chung, W. H., Hwa-Yaw, T., Wai, P. K. A. & Khandelwal, A. Time- and wavelength-division multiplexing of FBG sensors using a semiconductor optical amplifier in ring cavity configuration. *IEEE Photonics Technol. Lett.* **17**, 2709–2711. https://doi.org/10.1109/LPT.2005.859484 (2005).
36. Amorebieta, J. *et al.* Sensitivity-optimized strongly coupled multicore fiber-based thermometer. *Opt. Laser Technol.* **145**, 107532. https://doi.org/10.1016/j.optlastec.2021.107532 (2022).
37. Tang, Z. *et al.* Sensitivity optimization of symmetric multi-core fiber strain sensor based on mode-coupling theory. *Infrared Phys. Technol.* **111**, 103517. https://doi.org/10.1016/j.infrared.2020.103517 (2020).
38. Huang, W.-P. Coupled-mode theory for optical waveguides: An overview. *J. Opt. Soc. Am. A* **11**, 963–983. https://doi.org/10.1364/JOSAA.11.000963 (1994).
39. Hudgings, J., Molter, L. & Dutta, M. Design and modeling of passive optical switches and power dividers using non-planar coupled fiber arrays. *IEEE J. Quantum Electron.* **36**, 1438–1444. https://doi.org/10.1109/3.892564 (2000).
40. Snyder, A. W. Coupled-mode theory for optical fibers. *J. Opt. Soc. Am.* **62**, 1267–1277. https://doi.org/10.1364/JOSA.62.001267 (1972).
41. Xia, C. *et al.* Supermodes in coupled multi-core waveguide structures. *IEEE J. Sel. Top. Quantum Electron.* **22**, 196–207. https://doi.org/10.1109/JSTQE.2015.2479158 (2016).
42. Villatoro, J. *et al.* Ultrasensitive vector bending sensor based on multicore optical fiber. *Opt. Lett.* **41**, 832–835. https://doi.org/10.1364/OL.41.000832 (2016).
43. Yin, G., Zhang, F., Xu, B., He, J. & Wang, Y. Intensity-modulated bend sensor by using a twin core fiber: Theoretical and experimental studies. *Opt. Express* **28**, 14850–14858 (2020).
44. Shrivastava, A. & Gupta, V. Methods for the determination of limit of detection and limit of quantitation of the analytical methods. *Chron. Young Sci.* **2**, 21 (2011).

## Acknowledgements

This work was supported in part by the European Regional Development Fund, in part by the Ministerio de Economía y Competitividad under projects TEC2015-638263-C03-1-R and PGC2018-101997-B-I00, in part by Ministerio de Ciencia e Innovación: under projects PID2021-122505OB-C31 and TED2021-129959B-C21, in part by the Gobierno Vasco/Eusko Jaurlaritza under projects IT1452-22 and ELKARTEK (KK 2021/00082 and KK 2021/00092), in part by the Swedish Science Council, Office of Naval Research Global (Award N62909-20-1-2033) and in part by Vinnova Innovair: Forskningsprojekt inom flygteknik (D.N. 2020-00187). The work of Josu

Amorebieta is funded by a PhD fellowship from the University of the Basque Country UPV/EHU. The authors would like to thank Kenny Hey Tow, Erik Zetterlund and Fredrik Laurell for useful discussions and support.

### Author contributions
The original draft of the paper was written by J.A. and W.M., and reviewed by J.P., G.D., C.F., A.O.-G., J.Z. and J.V. J.A. and A.O.-G. collaborated in the theoretical approach, J.A., W.M., J.P. and C.F. designed and performed the experiments, and processed and analyzed data. A.O.-G. did the simulations, W.M. supervised the experiments. All authors discussed the experimental data, revised and approved the manuscript. J.A. and W.M. wrote the final version with inputs of all the authors.

### Competing interests
The authors declare no competing interests.

### Additional information
**Correspondence** and requests for materials should be addressed to J.A.

**Reprints and permissions information** is available at www.nature.com/reprints.

**Publisher's note**  Springer Nature remains neutral with regard to jurisdictional claims in published maps and institutional affiliations.

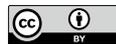